\documentclass[conference]{IEEEtran}
\usepackage{graphicx, epstopdf}
\usepackage{epsfig,cite}
\usepackage{color}
\usepackage{amssymb,amsmath,mathtools, mathrsfs}
\usepackage{enumitem}
\usepackage{graphicx}
\usepackage{epstopdf}
\usepackage{caption}
\usepackage[version=3]{mhchem}
\definecolor{blue}{rgb}{0,0,1}
\definecolor{darkgreen}{rgb}{0,.5,0}
\definecolor{darkred}{rgb}{.5,0,0}

\def\kth{$k^{\text{th}}$}
\def\ith{$i^{\text{th}}$}

\renewcommand{\vec}[1]{\mathbf{#1}}

\newcommand{\realSet}{\mathcal{R}}

\newcommand{\norm}[1]{\left\lVert#1\right\rVert}
\newcommand{\KLD}[2]{D_{\text{KL}}\left( #1\parallel#2\right)}

\def\sgn{\mathop{\rm sgn}\nolimits} 

\IEEEoverridecommandlockouts

\begin{document}
\bstctlcite{ICC09_Ref2:BSTcontrol}

\title{A Novel Experimental Platform for In-Vessel Multi-Chemical Molecular Communications}

\author{ \IEEEauthorblockN{Nariman~Farsad, David Pan, 
		and Andrea Goldsmith} 
		\IEEEauthorblockA{Department of Electrical Engineering, Stanford University, Stanford, CA, USA}
		\vspace{-0.7cm}

%
}

%
\maketitle

\begin{abstract}
This work presents a new multi-chemical experimental platform for molecular communication where the transmitter can release different chemicals. This platform is designed to be inexpensive and accessible, and it can be expanded to simulate different environments including the cardiovascular system and complex network of pipes in industrial complexes and city infrastructures. To demonstrate the capabilities of the platform, we implement a time-slotted binary communication system where a bit-0 is represented by an acid pulse, a bit-1 by a base pulse, and information is carried via pH signals. The channel model for this system, which is nonlinear and has long memories, is unknown. Therefore, we devise novel detection algorithms that use techniques from machine learning and deep learning to train a maximum-likelihood detector. Using these algorithms the bit error rate improves by an order of magnitude relative to the approach used in previous works. Moreover, our system achieves a data rate that is an order of magnitude higher than any of the previous molecular communication platforms.   
\end{abstract}


\vspace{-0.15cm}
\section{Introduction}
\vspace{-0.15cm}

In molecular communication, chemical signals are used to transfer information at various length scales from micro and nano meters to a few meters \cite{eckBook,far16ST}. Information can be modulated on different properties of information particles that are released by the transmitter.  These include the concentration/number \cite{pie10,kur12}, the type \cite{kim13}, or the time of release \cite{sri12}. Information particles can be transported from the transmitter to the receiver using diffusion, active transport, bacteria, and flow (see \cite{far16ST} and the references therein).

One of the main challenges in molecular communication is the lack of experimental data to validate the models. 
To bridge this gap between theory and practice, several previous works have focused on building experimental platforms for molecular communication. In \cite{far13}, a molecular communication platform is developed using a electronically controllable spray for transmission, aerosolized alcohol for carrying information, and a metal-oxide sensor, used in breathalyzers, for detection at the receiver. It was shown that data rates of 0.3 bits per second (bps) could be obtained on this platform. Using experimental data, it was demonstrated that this system's response does not match the accepted mathematical models, and corrections to the models were proposed \cite{far14JSAC}. This platform was expanded into a multiple-input multiple-output (MIMO) setup, where it was shown that the data rate can increase using MIMO \cite{koo16}.

In this work, we focus on molecular communication systems where there are {\em multiple chemical interactions} in the environment, and present a novel experimental platform where the transmitter can release different types of chemicals. Although it is difficult to find analytical models for these systems \cite{cho13NanoBio}, the inherent nonlinearities can result in interesting effects such as pattern formation and chemical oscillator \cite{reactDiffBook}. Moreover, multiple chemicals can be used to reduce the intersymbol interference (ISI) as was shown in \cite{far16SPAWC}. 

Our platform uses peristaltic pumps to inject different chemicals into a main fluid flow in a small silicon tube. The tubes can be networked in branches to replicate a more complex environment such as the cardiovascular system in the body or complex networks of pipes in industrial complexes and city infrastructures. This platform is deliberately designed to be inexpensive and easy to build so that other researchers can replicate it without having access to wet labs. 
Our platform can be used to by-pass computer simulations, which can be computationally-complex, by providing actual measurements relatively quickly. It can also be automated to run many experiments with minimum supervision.

To test our platform, we use acids and bases as transmission signals, and the pH value as the information bearing signal. As was discussed in \cite{far16SPAWC}, using this setup, the pH signal can both be {\em decreased} or {\em increased} through transmitting acids or bases. This resembles having negative and positive signals, which is something that was lacking in concentration-based molecular communication. Since analytical models for this system are nonexistent, a natural question to ask is the following: how can we do detection given the system's {\em nonlinearity} and {\em long memory}? Note that both of these characteristics of this system makes detection very difficult. Since an acid transmission would decrease the pH level in the environment and a base pulse would increase the pH level, one approach is to use the rate of change of pH. However, it is not clear if this approach, which we refer to as the {\em slope detector}, is optimal. 
\begin{figure*}[t]
	\begin{center}
		\includegraphics[width=0.9\textwidth,keepaspectratio]{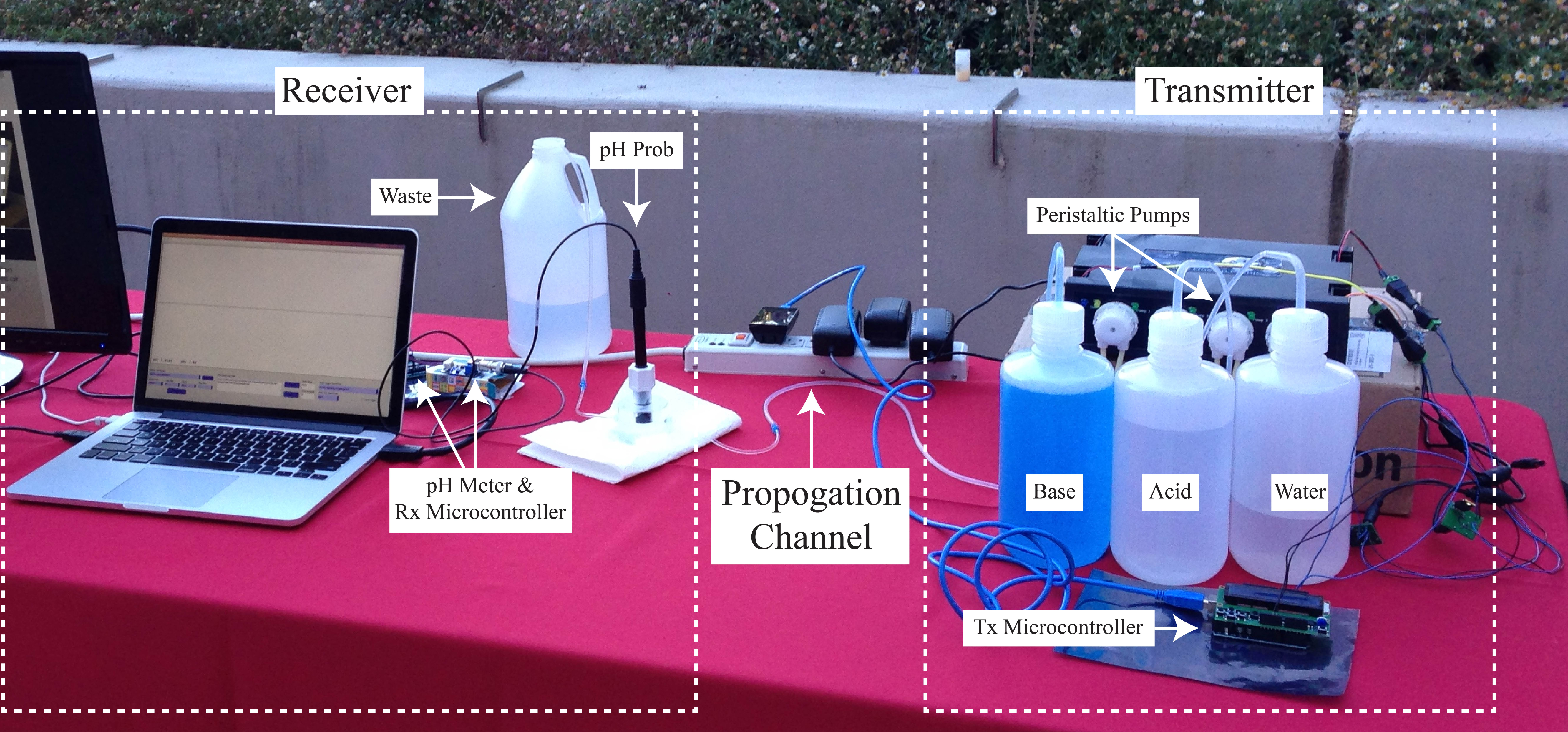}
	\end{center}
	\vspace{-0.4cm}
	\caption{\label{fig:thePlatform} The multi-chemical molecular communication platform.
		\vspace{-0.6cm}}
\end{figure*}

One solution to this problem is using detection algorithms inspired by machine learning \cite{far17Patent}. In this work we present two such algorithms: one based on support vector machines (SVM) \cite{smo04}, and one based on recurrent neural networks (RNN) \cite{lip15}.  In both cases, the detection algorithms have to go through a training phase based on known transmission sequences to ``learn'' the parameters of detection. At first glance, this may seem like extra overhead. However, if we consider the fact that there are no known models for multi-chemical communication channels, experimentation to validate any model would become inevitable. In fact, even in wireless radio communication many experimental data are used to ``train'' and ``learn'' (i.e., validate and revise) the models. This results in standardized models such as the Rayleigh fading channel and the corresponding detection algorithms. Note that the detection algorithms we present in this work can be used as part of {\em any communication system} (not necessarily molecular). They are especially useful when the underlying analytical channel model is unknown.     

Based on experimental data collected on our platform, we show that both the SVM based detector and the RNN based detector perform significantly better than the slope detection approach. We also demonstrate that the RNN detector behaves similar to a maximum likelihood detector, which maximizes the likelihood of the current symbol given the previously received signals and the current signal. Therefore, the RNN learns to do detection and equalization. This property is validated using experimental data, where for short symbol intervals, which result in severe ISI, the RNN's bit-error rate (BER) is about 10 times lower than the slope detection approach and about two times lower than the SVM detector. We demonstrate that data rates up to 4 bps are achievable on this platform, which is about 10 times better than any previous works \cite{far13,koo16}. 


The rest of this paper is organized as follows. In Section \ref{sec:expPlatform}, we present our multi-chemical communication platform. Then the modulation scheme, the communication protocols, and the detection algorithms necessary to implement a communication system on our platform are discussed in Section \ref{sec:implementation}. Numerical BER evaluations of the proposed detection algorithms are presented in Section \ref{sec:results}, and Section \ref{sec:conclusion} concludes the paper.

\vspace{-0.10cm}
\section{Multi-Chemical Experimental Platform}
\label{sec:expPlatform}
\vspace{-0.10cm}

In this section, we describe the construction of our experimental platform. To decide what type of chemicals are released by the transmitter, we need to decide what type of sensor(s) will be available at the receiver. In this paper, we assume that the information carrying particles are hydrogen ions and we use a pH sensor to detect their concentration. The transmitter can release acids and bases to decrease and increase the pH.

Note that although we consider this specific setup, the same general platform can be used with other types of chemicals. For example, instead of a pH sensor, a sensor that can detect glucose or proteins can be used. The transmitter can then release any number of chemicals including glucose, proteins or enzymes that degrade them. We now describe the transmitter, the propagation channel, and the receiver.

\vspace{-0.15cm}
\subsection{The Transmitter}
\vspace{-0.15cm}
The transmitter consists of a set of peristaltic pumps that can be used to pump different liquids. Peristaltic pumps use rotary motors attached to ``rollers'' to compress a flexible tube and generate flow. As the rotary motor turns, the part of the tube under the rollers is pinched and almost closed, which forces the liquid in the tube to move and generate flow as the rollers turn. This process, which is called peristalsis, is used in many biological systems such as the gastrointestinal system in the body. 

Peristaltic pumps are used in a wide variety of industries such as food, agriculture, medicine, engineering, and water and waste management. For example, they are used in dialysis machines to pump the blood in and out of the body and help patients with kidney failure. The peristaltic pumps we chose for our setup are aquarium dosing pumps, which can be purchased for as low as \$15 per pump. We use a custom metal-oxide semiconductor field-effect transistor (MOSFET) switch board to control the pumps through an Arduino Uno microcontroller.

The transmitter also has containers for the chemicals that can be released by the transmitter. Fig.~\ref{fig:thePlatform} shows a transmitter that can release acids and bases. Note that this architecture can be easily extended to use more chemicals by adding extra pumps and containers.

\vspace{-0.15cm}
\subsection{The Propagation Channel}
\vspace{-0.1cm}
The propagation channel is a silicon tube with a diameter of 2.5 mm. We use these small tubings since they are inexpensive and can be easily connected together to form complex networks that may resemble the cardiovascular system in the body. Fig.~\ref{fig:thePlatform} shows the propagation channel through a single tube. The single tube can resembles a vein in the body.

The liquid that constantly flows in the tube is filtered water. The flow is generated by a peristaltic pump that is always on. The volumetric flow generated by the pump has been  measured to be about 1.3 mL/s. Interestingly, the pH of the water has been measured to vary from pH 7 to pH 8.5 during different months. Note that instead of water, any other liquid can be easily substituted to be the main liquid flow in the channel. For example, it is possible to pump blood instead of water. 

The acid and the base are injected into the main water flow by the transmitter. The water flow would then carry these chemical signals to the receiver (i.e., the pH probe) and then a waste container. The system can be modified such that the flow would circulate back instead of being removed to a waste container. Note that such a closed loop system can be a more accurate representation of an in-body communication system. One benefit of using chemicals that react and cancel out, such as acids and bases, is that in closed loops the average concentration of these chemicals (e.g., pH) would remain constant. This is especially of interest for in-body communication.  

\vspace{-0.15cm}
\subsection{The Receiver}
\vspace{-0.1cm}
The receiver in our multi-chemical platform may have multiple detectors. However, for the acid-base system, information is encoded in the pH values. Therefore, a pH probe is used, where the pH value is converted into an electric potential. Particularly, the pH probe is connected to a pH meter, which is connected to an Arduino Uno microcontroller. The 10 bit analog-to-digital converter (ADC) on the Arduino is used to digitize the pH readings. The Arduino is then connected to a computer for display and potential processing and detection. Fig.~\ref{fig:thePlatform} shows the receiver and all its components.

One of the benefits of using a pH probe is that the relationship between the output electric potential and the pH level is linear under normal conditions (i.e., when the probe has not aged or is not damaged). Particularly, for the neutral pH, the output of the probe is zero volts, for the basic solutions, the output is negative voltage, and for acidic solutions, it is a positive voltage. Therefore, we can have positive and negative signals.  This is because the positive and negative ``chemical signals'' (i.e., acids and bases) can cancel out just like positive and negative numbers. For example, an acidic signal with pH 4 would cancel out a basic signal of pH 10. This property can be beneficial in many applications by allowing us to create different chemical patterns at the receiver and potentially create {\em chemical beamforming}.



In the next section, we describe how a communication system can be implemented on this platform. 

\vspace{-0.15cm}
\section{Implementing a Communication \\ System On the Platform}
\label{sec:implementation}
\vspace{-0.1cm}
The communication implementation can be divided into two parts. In the first part we describe the modulation scheme and the communication protocols, and in the second part we describe the detection process at the receiver.

\vspace{-0.15cm}
\subsection{Modulation and Protocol}
\label{sec:modulation}
\vspace{-0.1cm}
Time-slotted communication is employed where the transmitter modulates information on acid and base signals by injecting these chemicals into the channel during each symbol duration. We use a simple binary modulation in this work where the bit-0 is transmitted by pumping acid into the environment for 30 ms at the beginning of the symbol interval, and bit-1 is represented by pumping base into the environment for 30 ms at the beginning of the symbol interval. The symbol interval then depends on the period of silence that follows the 30 ms chemical injection. In particular, three different pause durations of 220 ms, 304 ms, 350 ms, and 470 ms are used in this work to represent bit rates of 4, 3, 2.6, and 2 bps.

\begin{figure}
	\begin{center}
		\includegraphics[width=0.9\columnwidth,keepaspectratio]{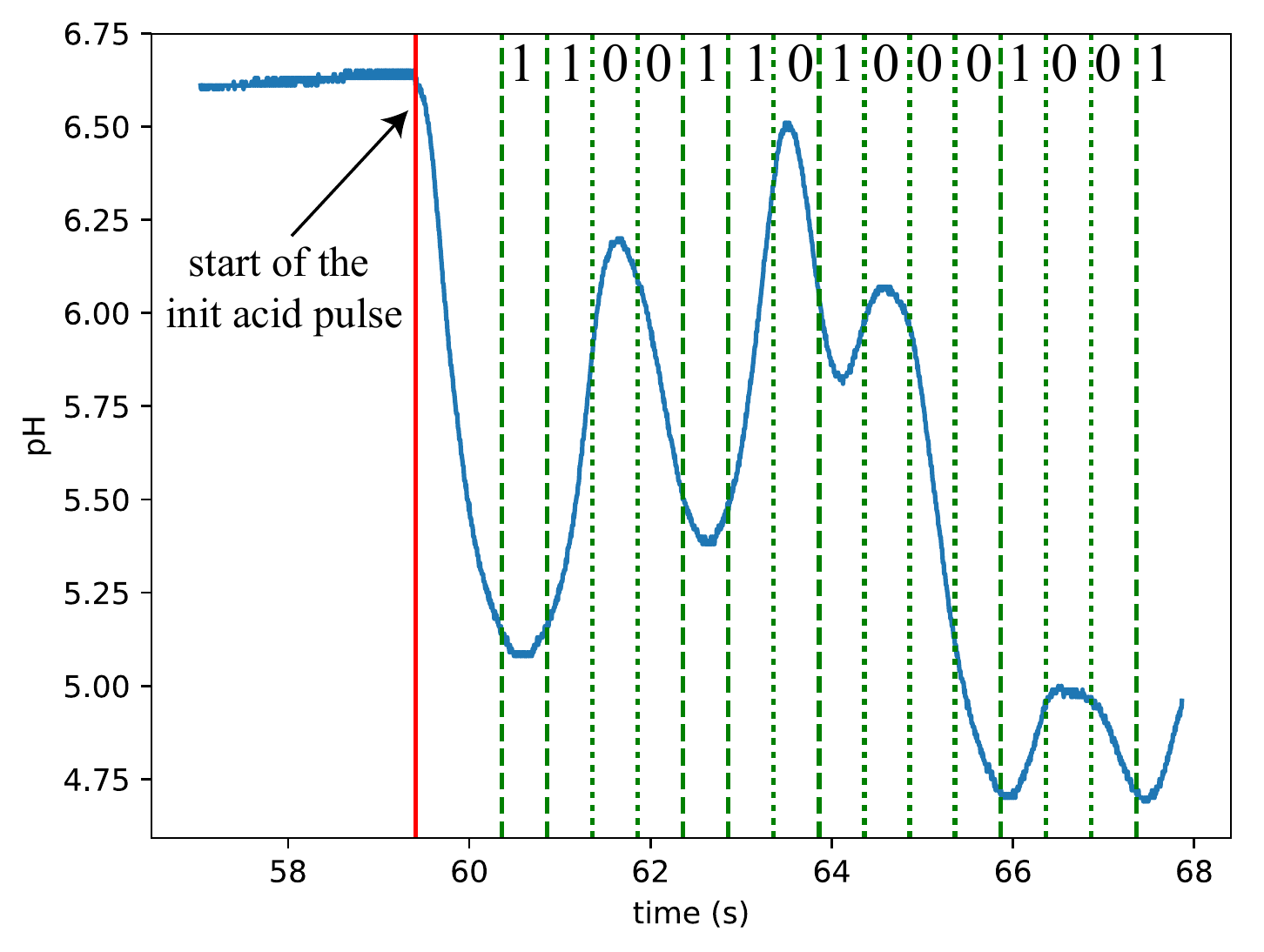}
	\end{center}
	\vspace{-0.4cm}
	\caption{\label{fig:sampleTrans} The received pH signal for the transmission of the bit sequence 110011010001001.}
	\vspace{-0.5cm}
\end{figure}
To synchronize the transmitter and the receiver, every message sequence starts with one initial injection of acid into the environment for 100 ms followed by 900 ms of silence. The receiver then detects the starting point of this pulse and uses it to synchronize itself with the transmitter. Fig.~\ref{fig:sampleTrans} shows the received pH signal for the transmission sequence ``110011010001001''. The start of the initial acid pulse detected by the receiver is shown using the red line. This detected time is used for synchronization and all the subsequent symbol intervals are shown by the green dashed and dotted lines. The dashed lines are used to indicate a bit-1 transmission and dotted lines to indicate a bit-0. A special termination sequence can be used to indicate the end of transmission. For example, if 5-bit encoded letters are transmitted \cite{far13}, the all zero sequence can be used to signal the end of transmission.  

\begin{figure*}
	\normalsize
	\centering
	\begin{minipage}{.43\textwidth}
		\begin{center}
			\includegraphics[width=\columnwidth,keepaspectratio]{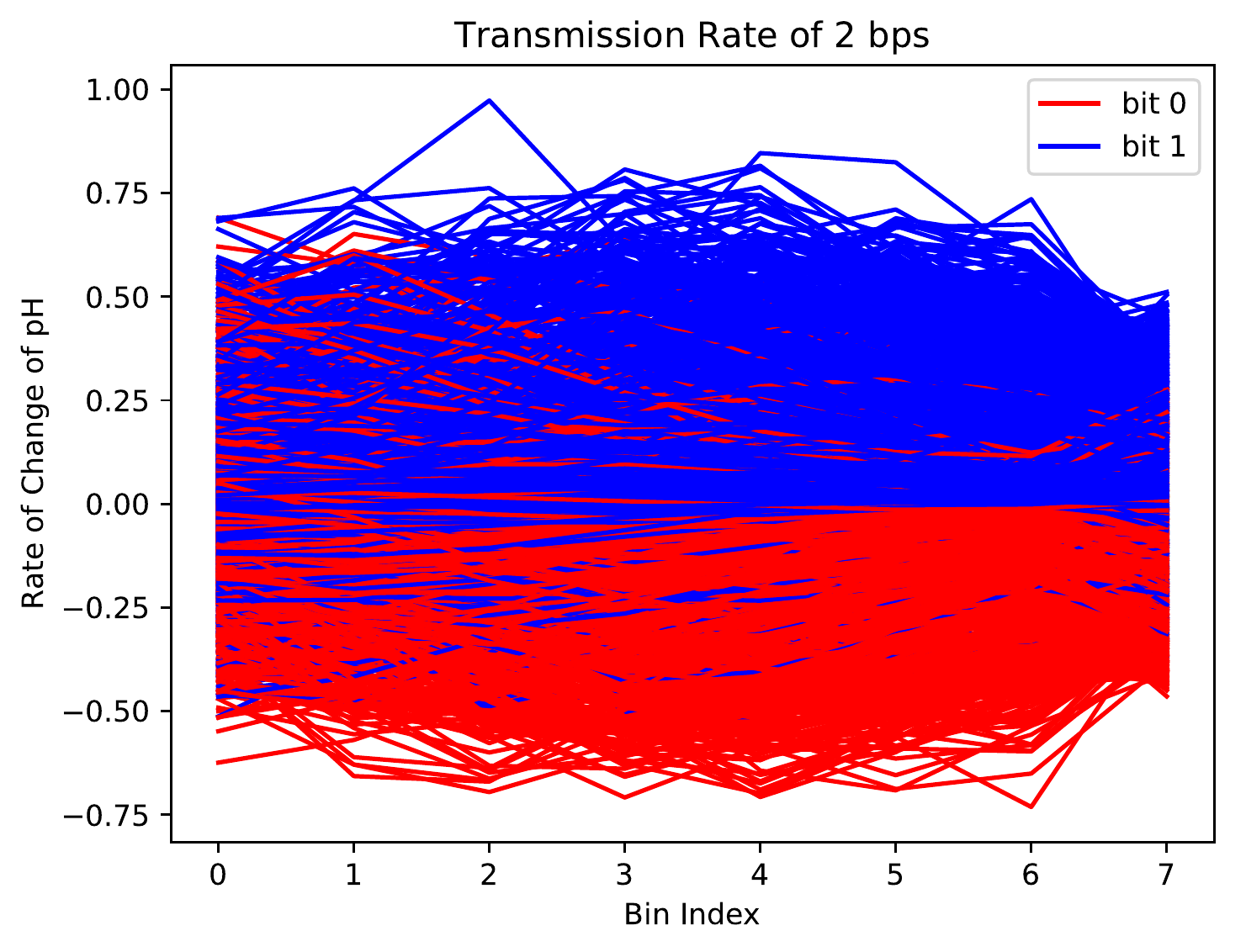}
		\end{center}
		\vspace{-0.4cm}
		\caption{\label{fig:RCeye2bps} Rate of change eye diagram at 2 bps.}
		\vspace{-0.5cm}
	\end{minipage}
	\hspace{0.6cm}
	\begin{minipage}{.43\textwidth}
		\begin{center}
			\includegraphics[width=\columnwidth,keepaspectratio]{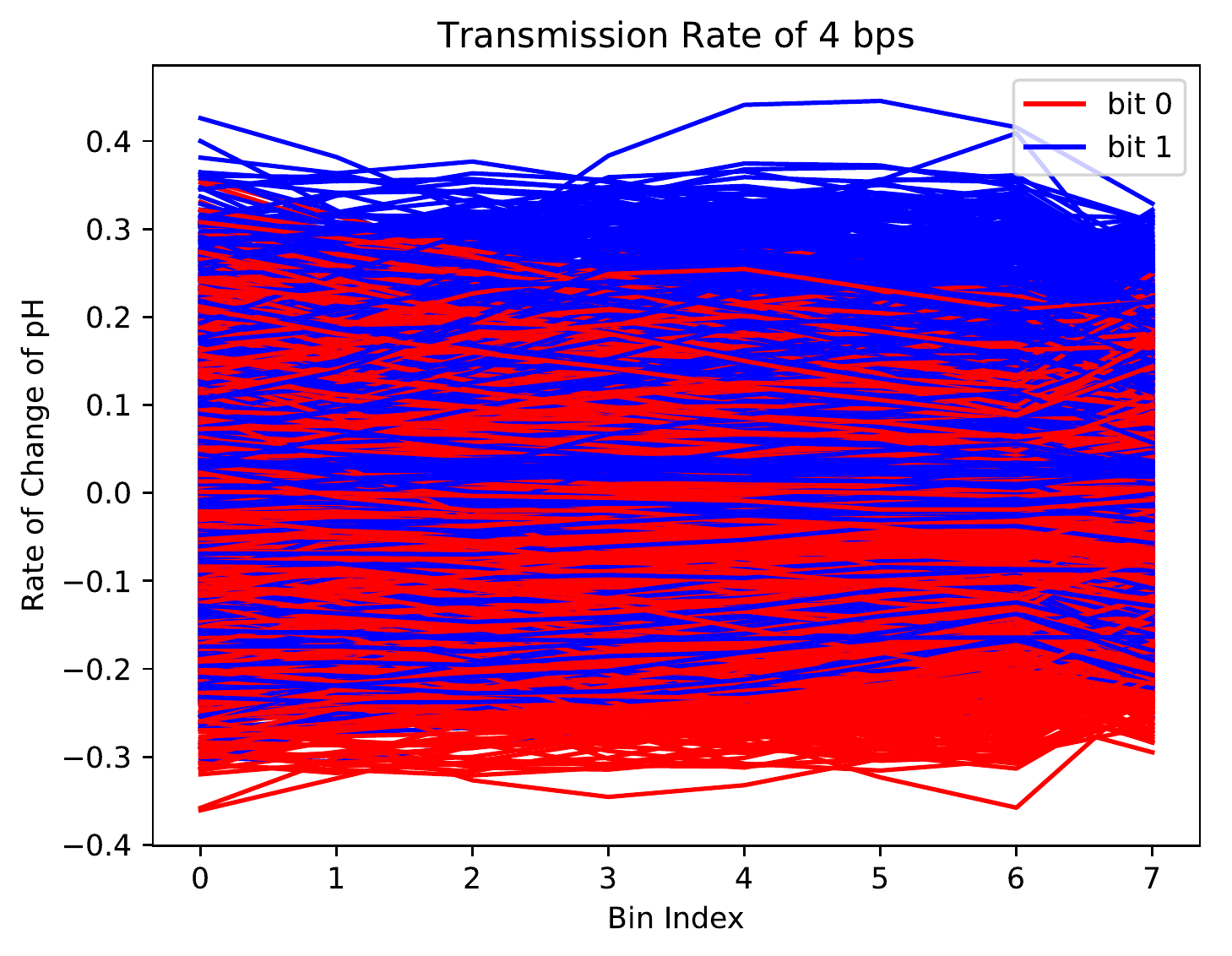}
		\end{center}
		\vspace{-0.4cm}
		\caption{\label{fig:RCeye4bps} Rate of change eye diagram at 4 bps.}
		\vspace{-0.5cm}
	\end{minipage}
\end{figure*}

\vspace{-0.15cm}
\subsection{Symbol Detection Based on Rate of Change}
\label{sec:slopeDetect}
\vspace{-0.1cm}
To design a detection algorithm for a communication system, the traditional approach is to develop analytical models for signal transmission, propagation and reception processes and then use these models to develop optimal or good detection algorithms. It is difficult, however, to obtain analytical models for multi-chemical communication systems, even when the average system behavior is considered instead of its stochastic nature \cite{reactDiffBook}. This is especially difficult when the chemicals react, which may result in nonlinearities and long system memory. Furthermore, it is difficult to model the noise introduced by the pumps and the sensor. Therefore, it is not immediately clear how to design detection algorithms for multi-chemical molecular communication systems. 

For our acid-base system, it is expected that when an acid pulse is transmitted, the pH should drop, and when a base pulse is injected into the environment, the pH should increase. Therefore, one approach to signal detection is to use the rate of change of pH to detect the symbols. To remove the noise from the raw pH signal, we divide the symbol interval into 8 bins by dividing the time into 8 equal subintervals. Then the pH values inside each bin are averaged to represent the pH value in the corresponding bit. Taking the difference of the pH of each bin would result in 7 rates of change of pH.

Figs.~\ref{fig:RCeye2bps} and \ref{fig:RCeye4bps} show the ``eye diagram'' of the rate of change of pH at 2 bps and 4 bps for many randomly transmitted bit sequences. As can be seen from the plots, the 7th rate of change is the best metric for detection of bits using a threshold based approach. Furthermore, as the data rate increases from 2 bps to 4 bps, the ISI increases, and it would be more difficult to detect the bits using this approach. This begs the question: can we do any better in terms of signal detection for our system given that we lack an analytical model for the channel?

\subsection{Symbol Detection Using Machine Learning}             
When analytical channel models are unknown, one approach to detection is to train a detector using machine learning \cite{far17Patent}. In this section, we describe two detection mechanisms based on machine learning: one based on support vector machines (SVM) \cite{smo04} for symbol-by-symbol detection, and one using recurrent neural networks (RNN) \cite{lip15} for detection of the current symbol based on the current and previous received signals. 

Let $x$ be the transmitted bit, and let $\vec{s}$ be a $d$-dimensional received signal vector. Let $\vec{y}$ be a length-$m$ feature vector that is extracted from the vector of received signal samples using $\varphi: \realSet^d \mapsto \realSet^m$. Note that the features could potentially be the received signal itself. 

A detection algorithm is trained by using a known sequence of transmissions and the corresponding features extracted from the received signals. We use two different techniques: one based on symbol-by-symbol detection, which detects each symbol independent of the other symbols, and one that considers ISI and uses the features from the previous symbols as well as the features from the current symbol for detection. In the rest of this section, we use the term {\em received features} to indicate the features extracted from the received signals during the symbol interval.

\subsubsection{Support Vector Machine Detector}
The first detector is trained using SVM \cite{smo04}. To understand SVM, we first consider linear regression where a linear relationship is used to estimate the transmitted symbol from the received features as follows:
\begin{align}
	\label{eq:linReg}
	\hat{x} = \vec{w}^{\top} \vec{y}+b,
\end{align}
where $\hat{x}$ is the estimated symbol, $\vec{w}$ are the weights, and $b$ the bias. To find the optimal weights and bias, a set of training data $\{(x_1,\vec{y}_1),(x_2,\vec{y}_2),\cdots,(x_n,\vec{y}_n) \}$ with $n$ samples of known transmission bits and corresponding received features are used. Particularly, the weights and bias that minimize the mean square error in the training set are used for detection.

In multi-chemical communication systems the relationship between the received signal and the transmitted signal is nonlinear. Therefore using SVM, the feature space is first mapped into a different and typically larger dimensional space using a function $\phi(\vec{y})$. A {\em kernel function} between two samples is defined to be the dot product:       
\begin{align}
	k\big(\vec{y}_i,\vec{y}_j\big) = \phi\big(\vec{y}_i\big)^{\top}  \phi\big(\vec{y}_j\big).
\end{align}
Let $x^\prime_i = 2x_i-1$. Using the kernel function, the corresponding $\phi$ function, and $x^\prime_i$, \eqref{eq:linReg} is transformed into \cite{smo04}
\begin{align}
	\label{eq:svm}
	\hat{x}^\prime = \sgn (\vec{w}^{\top} \phi(\vec{y})+b) = \sgn \bigg( b+\sum_{i=1}^{n} \alpha_i  x^\prime_i k(\vec{y},\vec{y}_i) \bigg),
\end{align}
where $i$ is the index of each sample in the training dataset, and $\alpha_i$ is a weight associated with that sample. Therefore to detect the symbols, the training dataset is used directly in the decision process.

Using the kernel function and the training dataset, the optimal $\alpha_i$ wights and the bias can be found through convex optimization \cite{smo04}.  Note that in SVM, most $\alpha_i$ weights in \eqref{eq:svm} will be zero, which means only a small fraction of the training samples are actually used during detection. The term {\em support vector} stems from this property and is used to indicate the training samples with nonzero weight. 

We use the Gaussian kernel, which is also known as the radial basis function (RBF) kernel. This kernel is given by 
\begin{align}
\label{eq:rbfKer}
k(\vec{y}_i,\vec{y}_j) = \exp\left( - \frac{\norm{\vec{y}_i-\vec{y}_j}^2_2}{2\sigma^2}\right),
\end{align}
where $\sigma^2$ is the parameter of the model and can be optimized numerically using the training data. This kernel is employed since it uses the Euclidean distance between the features as a figure of merit, maps the feature space into an infinite dimensional space, and performs linear regression in that space.  

\subsubsection{Recurrent Neural Network Detector}
Although the SVM detector presented in the previous section can capture some of the nonlinearities of the communication system, it does not capture the ISI and the system memory. Therefore, in this section we propose a detector based on RNNs that captures these effects by training on sequential data \cite{lip15}. 


Let $\bar{\vec{x}}^{(i)} = P(x^{(i)})$ be the probability mass function (PMF) corresponding to the \ith~transmission in a sequence of $K$ consecutive transmissions. In the training dataset, since it is known exactly which symbol is transmitted, all the elements of $\bar{\vec{x}}^{(i)}$ are zero except the element corresponding to $x^{(i)}$ (which is equal to 1).  
Let $\vec{X}^{(K)} = [\bar{\vec{x}}^{(1)},\bar{\vec{x}}^{(2)},\cdots,\bar{\vec{x}}^{(K)}]$ be a sequence of consecutively transmitted symbols, and $\vec{Y}^{(K)} = [\vec{y}^{(1)},\vec{y}^{(2)},\cdots,\vec{y}^{(K)}]$ the corresponding sequence of received feature vectors. Note that instead of the symbols we are using their PMF indicator vector. In the machine learning literature this is known as the {\em one-hot} representation, where the location of the element which equals 1 indicates the transmitted symbol. To train the RNN, we use a training dataset $\{(\vec{X}^{(K)}_1,\vec{Y}^{(K)}_1),(\vec{X}^{(K)}_2,\vec{Y}^{(K)}_2), \cdots, (\vec{X}^{(K)}_n,\vec{Y}^{(K)}_n) \}$, which consists of $n$ samples of length-$K$ sequences.

In RNNs, a hidden layer is used to keep an internal state that depends on previous inputs to the system (i.e., the received features corresponding to previous symbol intervals). The current symbol is then estimated based on the internal state of the RNN and the features extracted from the current received signal. Let $\vec{h}^{(k)}$ be the hidden state of the RNN at the end of the \kth transmission. Then, a simple generic RNN architecture can be represented by \cite{lip15}
\begin{align}
	\vec{a}^{(k)} &= \vec{b} + \vec{W} \vec{h}^{(k-1)} + \vec{U} \vec{y}^{(k)},\\
	\vec{h}^{(k)} &= \tanh \big(\vec{a}^{(k)}\big), \\
	\vec{o}^{(k)} &= \vec{c} + \vec{V} \vec{h}^{(k)}, \\
	\hat{\vec{x}}^{(k)} &= \text{softmax}(\vec{o}^{(k)}), \label{eq:softmaxLayer}
\end{align}
where $\vec{b}$ and $\vec{c}$ are bias parameters of the RNN, and $\vec{W}, \vec{U},$ and $\vec{V}$ are the weight parameters of the RNN. The auxiliary variable $\vec{a}^{(k)}$ is updated based on the previous hidden state $\vec{h}^{(k-1)}$, and the feature vector $\vec{y}^{(k)}$ extracted from the signal observed during the current symbol interval. The hidden state and the estimated output are then updated based on $\vec{a}^{(k)}$.

The final softmax operation in \eqref{eq:softmaxLayer} converts the output into a PMF form where the output vector sums up to one \cite{lip15}. Particularly, because the hidden state of the RNN stores information about the received features from the previous symbol intervals, the output can be thought of as a conditional distribution:    
\begin{align}
	\label{eq:condiDistView}
	\hat{\vec{x}}^{(i)} = P(x^{(i)} \mid \vec{h}^{(i-1)}, \vec{y}^{(i)}) \approx P(x^{(i)} \mid \vec{Y}^{(i)}).
\end{align}
From this conditional PMF, the index of the element with the highest probability is selected as the transmitted symbol. To find the optimal weight and bias parameters of the RNN using the training dataset, a loss function $\mathcal{L}$ is defined using {\em cross entropy} as
\begin{align}
	\label{eq:crossEnt}
	\mathcal{L} = \sum_{i=1}^{K}H(\bar{\vec{x}}^{(i)},\hat{\vec{x}}^{(i)}) = \sum_{i=1}^{K}H(\bar{\vec{x}}^{(i)}) + \KLD{\bar{\vec{x}}^{(i)}}{\hat{\vec{x}}^{(i)}},
\end{align} 
where $H(\bar{\vec{x}}^{(i)},\hat{\vec{x}}^{(i)})$ is the cross entropy between the correct PMF and the estimated PMF by the RNN, and $\KLD{.}{.}$ is the Kullback-Leibler divergence. During the training the weights and biases of the RNN are adjusted iteratively using techniques such as the stochastic gradient descent algorithm to minimize this loss. Note that minimizing the loss is equivalent to minimizing the cross-entropy or the Kullback-Leibler divergence distance between the true PMF and the RNN estimated PMF.

Alternatively, using \eqref{eq:condiDistView} it can be shown that \eqref{eq:crossEnt} can be written as
\begin{align}
\label{eq:likelihood}
\mathcal{L} = -\sum_{i=1}^{K} \log\bigg(  \hat{\vec{x}}^{(i)} [x^{(i)}]\bigg)  = -\sum_{i=1}^{K} \log P_{\text{model}}(x^{(i)}\mid \vec{Y}^{(i)}),
\end{align} 
where $\hat{\vec{x}}^{(i)}[x^{(i)}]$ indicates the element of $\hat{\vec{x}}^{(i)}$ corresponding to symbol $x^{(i)}$, and $P_{\text{model}}(x^{(i)}\mid \vec{Y}^{(i)})$ is the probability of symbol $x^{(i)}$ given the received features from the current and previous symbol intervals, which is obtained by the RNN model. Note that minimizing \eqref{eq:likelihood} is equivalent to maximizing the log expression in the summations, which essentially maximizes the log-likelihoods. Therefore, during the training, known transmission data are used to train a detector that {\em maximizes log-likelihoods}.  The detector is in essence similar to a maximum-likelihood (ML) detector, where the structure of the neural network effects the complexity of the underlying model that relates the transmitted and the received signal, and the training data is used to tune the parameters of this model. 


Although here we have presented a generic RNN architecture due to lack of space, more complex networks and multi-layer architectures can be used in practice to capture a wide class of linear and nonlinear channel models. Particularly, in our platform we use the long short-term memory (LSTM) networks, which have more nodes compared to the generic model presented here and have a better training performance. In the next section, we compare the performance of all these detectors using experimental data.

\vspace{-0.15cm}
\section{Results}
\label{sec:results} 
\vspace{-0.15cm}
We use the experimental data collected on our platform to compare the performance of each detector. In particular, 194 experimental data are collected, where for each experiment, 120 random bits are transmitted using the acid-base modulation scheme discussed in Section \ref{sec:modulation}. This results in a total of more than 23 thousand transmitted bits. The symbol intervals in these experiments are 250 ms, 334 ms, 380 ms, and 500 ms, respectively, which results in data rates ranging from 4 bps to 2 bps.

From the data, 20 percent of the experiments with symbol intervals of 250 ms, 334 ms, and 380 ms are used as test datasets. The rest of the data is used for training and tuning the parameters of the detection algorithm. The first detection algorithm, which we refer to as the {\em baseline algorithm}, is the slope detector presented in Section \ref{sec:slopeDetect}. Using the training data we find that the rate of change in the last bin is best for detection, and the optimal decision threshold is 0.

The features used for the SVM detection algorithm are the 8 pH values corresponding to each bin, which was explained in Section \ref{sec:slopeDetect}, the mean and the variance of the 8 values, the 7 rates of change of pH between bins, and the mean and the variance of the 7 rates of changes. We found that this set of features resulted in the best detection performance. Using hyper parameter tuning, $\sigma^2=5$ yielded the best results for the RBF kernel in \eqref{eq:rbfKer}. For the RNN detector, we use a single layer LSTM cell with an internal state size of 16. We use the average pH value of the 8 bins and the 7 rate of pH changes as the input features to the RNN, which resulted in the best performance.      
\begin{figure}
	\begin{center}
		\includegraphics[width=1\columnwidth,keepaspectratio]{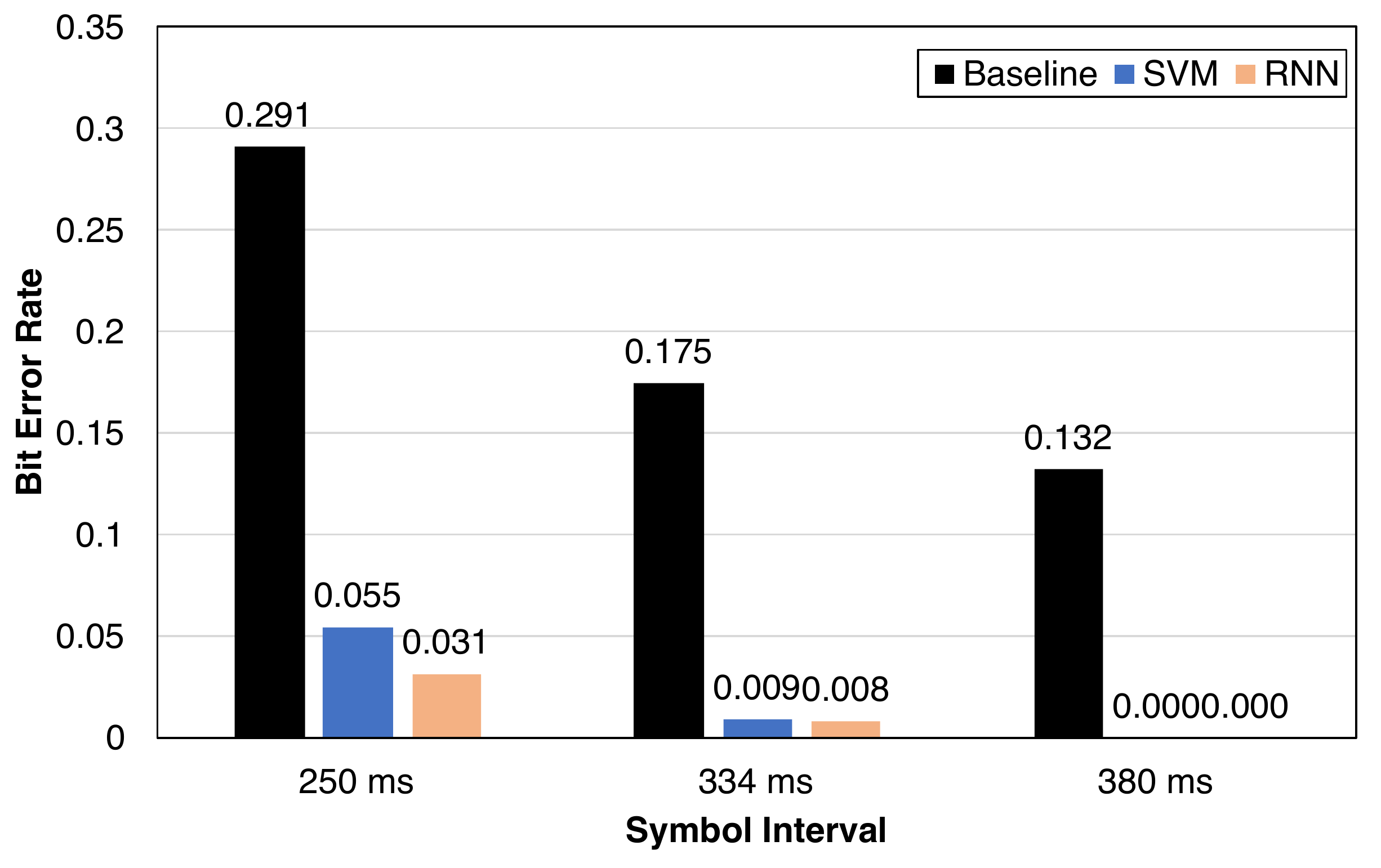}
	\end{center}
	\vspace{-0.35cm}
	\caption{\label{fig:detctBarPlot}Performance of the different detection algorithms.}
	\vspace{-0.5cm}
\end{figure}

Fig.~\ref{fig:detctBarPlot} shows the bit error rate (BER) of each detection algorithm for different symbol intervals. The number of bits in the test data for each symbol interval is 1080. the performance of all detectors improves with the symbol interval since the intersymbol interference is reduced. Both detection techniques based on machine learning significantly outperform the baseline slope detector. The RNN detector performs better than the SVM detector, especially at smaller symbol intervals (i.e., higher data rates), where ISI is severe. At 250 ms (4 bps), the RNN detector's BER is almost half that of the SVM detector and about 10 times better than the baseline detector. This is because the RNN detector captures the ISI in the channel by design and ``equalizes'' its effects. This demonstrates the significant impact machine learning techniques could have in designing molecular communication systems and more broadly in many other areas in communications.    

\vspace{-0.1cm}
\section{Conclusions}  
\label{sec:conclusion}
\vspace{-0.15cm}
We presented a new multi-chemical molecular communication platform for chemical communication inside small tubes. 
This system can be expanded to simulate different environments including the cardiovascular system and complex network of pipes in industrial complexes and city infrastructures. A chemical communication system was implemented on this platform where acids and bases are used as transmission chemicals, using the pH as the information carrying signals. Novel detection algorithms, inspired by machine learning and deep learning, were proposed and implemented on the system. These algorithms showed 10 times improvement in performance compared to a slope-based detection used in previous works, and the system was shown to achieve data rates that are 10 time better than any previous platforms.

\vspace{-0.15cm}
\bibliographystyle{IEEEtran}
\bibliography{IEEEabrv,MolCom_YearSorted}

\begin{thebibliography}{10}
\providecommand{\url}[1]{#1}
\csname url@samestyle\endcsname
\providecommand{\newblock}{\relax}
\providecommand{\bibinfo}[2]{#2}
\providecommand{\BIBentrySTDinterwordspacing}{\spaceskip=0pt\relax}
\providecommand{\BIBentryALTinterwordstretchfactor}{4}
\providecommand{\BIBentryALTinterwordspacing}{\spaceskip=\fontdimen2\font plus
\BIBentryALTinterwordstretchfactor\fontdimen3\font minus
  \fontdimen4\font\relax}
\providecommand{\BIBforeignlanguage}[2]{{%
\expandafter\ifx\csname l@#1\endcsname\relax
\typeout{** WARNING: IEEEtran.bst: No hyphenation pattern has been}%
\typeout{** loaded for the language `#1'. Using the pattern for}%
\typeout{** the default language instead.}%
\else
\language=\csname l@#1\endcsname
\fi
#2}}
\providecommand{\BIBdecl}{\relax}
\BIBdecl

\bibitem{eckBook}
T.~Nakano, A.~W. Eckford, and T.~Haraguchi, \emph{Molecular communication},
  1st~ed.\hskip 1em plus 0.5em minus 0.4em\relax Cambridge University Press,
  2013.

\bibitem{far16ST}
N.~Farsad, H.~B. Yilmaz, A.~Eckford, C.-B. Chae, and W.~Guo, ``A comprehensive
  survey of recent advancements in molecular communication,'' \emph{{IEEE}
  Commun. Surveys Tuts.}, vol.~18, no.~3, pp. 1887--1919, 2016.

\bibitem{pie10}
M.~Pierobon and I.~F. Akyildiz, ``A physical end-to-end model for molecular
  communication in nanonetworks,'' \emph{{IEEE} J. Sel. Areas Commun.},
  vol.~28, no.~4, pp. 602--611, May 2010.

\bibitem{kur12}
M.~S. Kuran, H.~B. Yilmaz, T.~Tugcu, and I.~F. Akyildiz, ``Interference effects
  on modulation techniques in diffusion based nanonetworks,'' \emph{Nano
  Commun. Netw.}, vol.~3, no.~1, pp. 65--73, Mar. 2012.

\bibitem{kim13}
N.-R. Kim and C.-B. Chae, ``Novel modulation techniques using isomers as
  messenger molecules for nano communication networks via diffusion,''
  \emph{{IEEE} J. Sel. Areas Commun.}, vol.~31, no.~12, pp. 847--856, Dec.
  2013.

\bibitem{sri12}
K.~V. Srinivas, A.~W. Eckford, and R.~S. Adve, ``Molecular communication in
  fluid media: The additive inverse {Gaussian} noise channel,'' \emph{{IEEE}
  Trans. Inf. Theory}, vol.~58, no.~7, pp. 4678--4692, July 2012.

\bibitem{far13}
N.~Farsad, W.~Guo, and A.~W. Eckford, ``Tabletop molecular communication: Text
  messages through chemical signals,'' \emph{PLoS ONE}, vol.~8, no.~12, p.
  e82935, Dec. 2013.

\bibitem{far14JSAC}
N.~Farsad, N.-R. Kim, A.~W. Eckford, and C.-B. Chae, ``Channel and noise models
  for nonlinear molecular communication systems,'' \emph{{IEEE} J. Sel. Areas
  Commun.}, vol.~32, no.~12, pp. 2392--2401, Dec 2014.

\bibitem{koo16}
B.~H. Koo, C.~Lee, H.~B. Yilmaz, N.~Farsad, A.~Eckford, and C.~B. Chae,
  ``Molecular mimo: From theory to prototype,'' \emph{{IEEE} J. Sel. Areas
  Commun.}, vol.~34, no.~3, pp. 600--614, March 2016.

\bibitem{cho13NanoBio}
C.~T. Chou, ``Extended master equation models for molecular communication
  networks,'' \emph{{IEEE} Trans. NanoBiosci.}, vol.~12, no.~2, pp. 79--92,
  June 2013.

\bibitem{reactDiffBook}
B.~Grzybowski, \emph{Chemistry in Motion: Reaction-Diffusion Systems for Micro-
  and Nanotechnology}.\hskip 1em plus 0.5em minus 0.4em\relax Wiley, 2009.

\bibitem{far16SPAWC}
N.~Farsad and A.~Goldsmith, ``A molecular communication system using acids,
  bases and hydrogen ions,'' in \emph{Proc. IEEE Workshop on Signal Process.
  Adv. in Wireless Commun. (SPAWC)}, July 2016, pp. 1--6.

\bibitem{far17Patent}
------, ``Systems and methods for transmitting data using machine learning
  classification,'' U.S. Provisional Patent 62/458,936, Feb.~14, 2017.

\bibitem{smo04}
A.~J. Smola and B.~Sch{\"o}lkopf, ``A tutorial on support vector regression,''
  \emph{Statistics and Computing}, vol.~14, no.~3, pp. 199--222, 2004.

\bibitem{lip15}
Z.~C. Lipton, J.~Berkowitz, and C.~Elkan, ``A critical review of recurrent
  neural networks for sequence learning,'' \emph{arXiv preprint
  arXiv:1506.00019}, 2015.

\end{thebibliography}

\end{document}